# Edge dependence of nonlocal transport in gapped bilayer graphene

Hyeon-Woo Jeong[1], Seong Jang[1], Sein Park[1], Kenji Watanabe[2], Takashi Taniguchi[3] and Gil-Ho Lee[1,*]

[1]Department of Physics, POSTECH, Pohang, Republic Korea

[2]Research Center for Functional Materials, National Institute for Materials Science, Tsukuba, Japan

[3]International Center for Materials Nanoarchitectonics, National Institute for Materials Science, Tsukuba, Japan

[*]Correspondence to Gil-Ho Lee (lghman@postech.ac.kr)

KEYWORDS. Bilayer graphene, tunable band gap, valley Hall effects, edge transport, etched boundary

ABSTRACT. The topological properties of gapped graphene have been explored for valleytronics applications. Prior transport experiments indicated their topological nature through large nonlocal resistance in Hall-bar devices, but the origin of this resistance was unclear. This study focused on dual-gate bilayer graphene (BLG) devices with naturally cleaved edges, examining how edge-etching with an oxygen plasma process affects electron transport. Before etching, local resistance at the charge neutral point increased exponentially with the displacement field and nonlocal resistance was well explained by ohmic contribution, which is typical of gapped BLG. After-etching, however, local resistance saturated with increasing displacement field, and nonlocal resistance deviated by three orders of magnitude from ohmic contribution. We suggest that these significant changes in local and nonlocal resistance arise from the formation of edge conducting pathways after the edge-etching, rather than from a topological property of gapped BLG that has been claimed in previous literatures.

Graphene presents novel electronic properties owing to the excitation of its low-energy Dirac band structure[1-5]. Bernal-stacked bilayer graphene (BLG) exhibits a massive Dirac spectrum with two pairs of low-energy parabolic bands[6]. The application of an external electric field breaks the inversion symmetry, that is, the layer symmetry of this spectrum, leading to the formation of a band gap at the Dirac point[6]. Although breaking the sublattice symmetry to open a gap in monolayer graphene is tricky[7, 8], BLG has an easily tunable band gap that can be manipulated by controlling the electric field using external gates[6, 9-11]. Band-gap opening in BLG using electric fields has been demonstrated through various experimental studies[9, 12-18], with the resultant materials showing potential applications in high-mobility transistors. In addition to conventional charge transport, the topological nature of graphene from the sublattice symmetry provides access to valley chiral transport[19-29]. The electrons in the valleys (K and K') of graphene with broken inversion symmetry have opposite Berry curvature signs, resulting in a valley Hall effect, wherein electrons in different valleys move in opposite directions. Therefore, the valley degree of freedom for electronics, called valleytronics[30, 31], resilience against bulk impurities and the ability to operate without the need for a magnetic field, must be considered. Previous studies have attempted to experimentally demonstrate valley polarization phenomena in gapped graphene systems. For instance, the Hall voltages in gapped graphene systems have been measured using illuminating circularly polarized mid-infrared light[32] or nonlocal current injection[21, 23, 24, 29], and the results have been widely used to claim the occurrence of valley polarized transport. Several studies on BLG devices suggest that the valley-polarized edge states result in a nonlocal resistance much larger than the value expected from ohmic contributions alone[23, 24]. However, the intervalley scattering near the atomically rough etched edges of BLG could have disrupted valley polarization phenomena[33], so the origin of this larger nonlocal resistance remains unclear. Furthermore, several researchers have reported that topologically trivial edge states could be formed by band-bending[34] or charge accumulation[33-36] at the edges, leading to a large nonlocal resistance.

In this study, we investigate how the formation of edges by reactive ion-etching affects the nonlocal resistance of gapped BLG. To avoid etching-induced edge roughness, we fabricated high-mobility BLG devices with preserved naturally cleaved edges. The areas of the top and bottom graphite gates were larger than that of the BLG layer to tune the carrier density and displacement field. This configuration ensured that the device was free from stray electric fields and etching-induced edge roughness. The nonlocal resistance of gapped BLG near the charge-neutrality point (CNP) was measured before and after edge-etching without affecting the bulk. Prior to edge-etching, the nonlocal resistance near the CNP was well-described by the ohmic contribution from the nonlocally injected bias current. After etching, however, the ratio of nonlocal resistance to ohmic contribution increased by up to two orders at the largest displacement field we have applied, while the local resistance stayed almost the same for the same bandgap. These changes in nonlocal and local resistance can be explained by the formation of edge conducting channels after etching, and not necessarily by the emergence of an opposite Berry curvature near the Dirac point for different valleys, as claimed in previous studies[23, 24].

RESULTS AND DISCUSSION

We investigated electron transport in the dual-gated BLG devices before and after the etching of the edges of the BLG layer. Rectangular BLG flakes were obtained using a standard mechanical exfoliation method. To minimize the introduction of impurities into BLG during its fabrication, we encapsulated the top and bottom of a BLG layer with hexagonal boron nitride (hBN) layers and adopted top and bottom graphite gates using the dry transfer technique[37, 38]. The bottom and top graphite gate layers completely covered the BLG region, thereby avoiding the electrostatic accumulation of carriers near the edge owing to the non-uniform electric field. Gold electrodes were attached to only the BLG layer, without touching the top and bottom graphite gates, by carefully etching a portion of the top graphite layer and insulating the edge of the bottom graphite layer (Figs. 1a, c). Etching of the BLG layer was avoided during the fabrication process (see details in the Methods section).

We tuned the displacement field, which is defined[9] as $D = (D_{\text{TG}} + D_{\text{BG}})/2$, to break the inversion symmetry of BLG and open its band gap $E_{\text{g}}$ (Fig. 1A) by controlling the top ($V_{\text{TG}}$) and bottom gate voltages ($V_{\text{BG}}$). The displacement fields from the top ($D_{\text{TG}}$) and bottom ($D_{\text{BG}}$) gates are given as $D_{\text{TG}} = -\epsilon_{\text{hBN}}(V_{\text{TG}} - V_{\text{T0}})/d_{\text{TG}}$ and $D_{\text{BG}} = \epsilon_{\text{hBN}}(V_{\text{BG}} - V_{\text{B0}})/d_{\text{BG}}$, respectively, where $\epsilon_{\text{hBN}}$ is the dielectric constant of h-BN, $d_{\text{TG}}$ and $d_{\text{BG}}$ refer to the thicknesses of the top and bottom h-BN layers, respectively, and $V_{\text{T0}}$ and $V_{\text{B0}}$ refer to the voltages applied to the top and back gates to bring BLG to the CNP, respectively. In this study, $\epsilon_{\text{hBN}}$, $d_{\text{TG}}$, $d_{\text{BG}}$, $V_{\text{T0}}$, and $V_{\text{B0}}$ were 3.9, 60 nm, 62 nm, −0.12 V, and 0.13 V, respectively. The total carrier density was also tuned as $n = \epsilon_0(D_{\text{BG}} - D_{\text{TG}})/e$, where $e$ and $\epsilon_0$ are the elementary charge and vacuum permittivity, respectively.

Figure 1d shows the local resistance $R_{\text{loc}}$ ($= R_{\text{SD},23}$) of device 1 measured as a function of $n$ at a fixed $D$ of $0.6\ V/nm$ before and after edge-etching. Here, $R_{\text{ij,kl}}$ represents the resistance, which is the voltage difference between terminals $k$ and $l$ divided by the bias current from terminals $i$ to $j$. For a highly doped regime ($|n| > 0.5 \times 10^{11}\ \text{cm}^{-2}$), little difference in $R_{\text{loc}}$ is observed before and after etching, as the highly conducting BLG bulk dominates the transport process. However, at the CNP, where bulk conduction is suppressed the most, $R_{\text{loc}}$ decreases by a factor of 4 after edge-etching (Figure 1d). This result suggests the formation of additional conduction channels near the etched edges because the BLG bulk is protected by the hBN and graphite layers and remains highly insulating at the CNP.

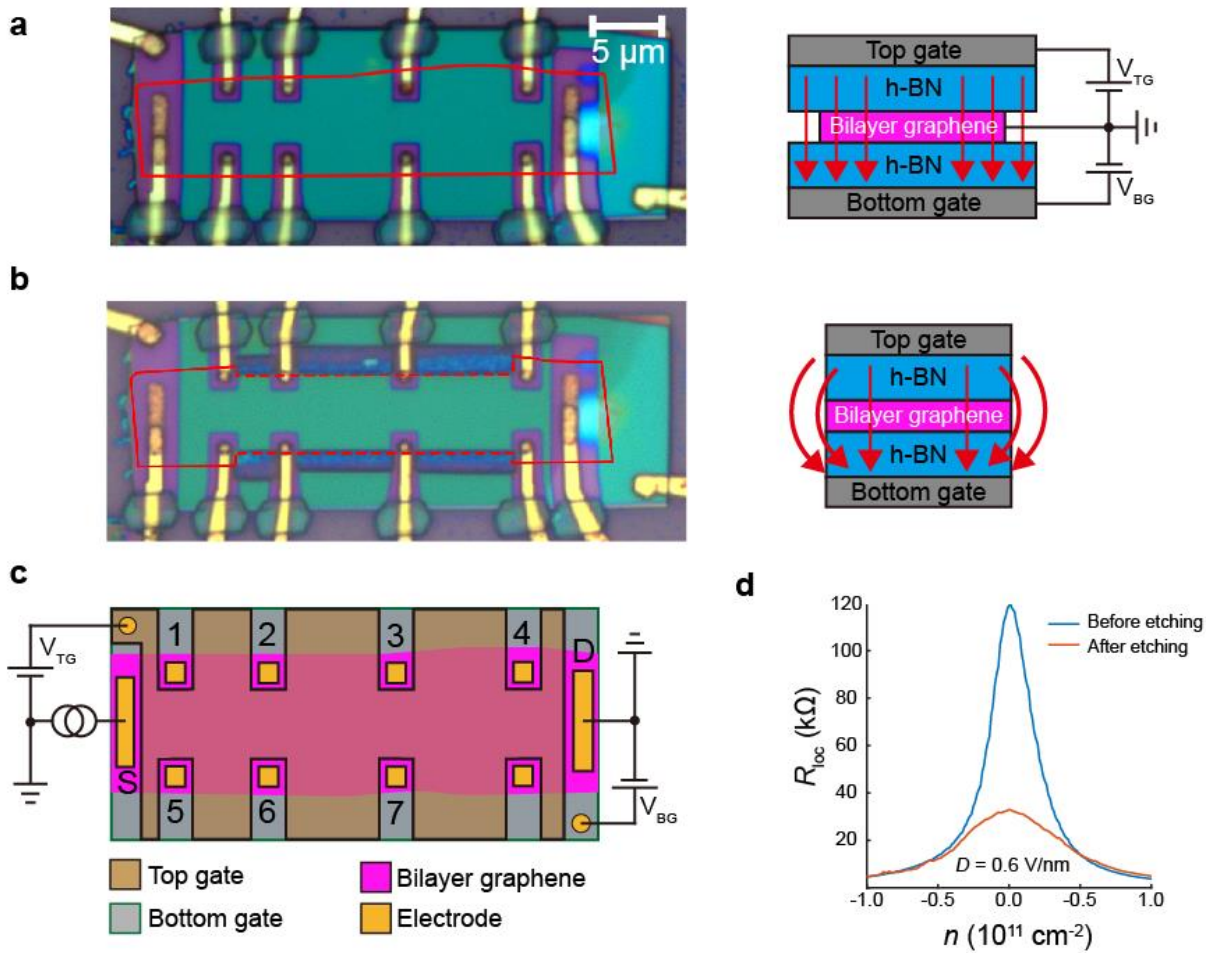

**Figure 1.** Local resistance of a dual-gated bilayer graphene (BLG) device (device 1) before and after edge-etching. (a, b) Optical microscope images and side-view schematics of the dual-gated BLG device (a) before and (b) after edge-etching. The naturally cleaved edge of BLG is outlined by the red solid line, while the etched edge is outlined by the red dashed line. The red arrows in the schematics represent the electric field induced by the dual gates. (c) Top-view schematics of the device before edge-etching. (d) Carrier density ($n$) dependence of local resistance ($R_{\text{loc}}$) measured at a fixed displacement field $D$ of 0.6 V/nm and temperature of 50 K before (blue) and after (red) edge-etching.

To investigate these additional conduction channels in more detail, we measured the local and nonlocal resistances $R_{\text{nl}}$ ($= R_{26,37}$) under varying $n$ and $D$ before and after edge-etching. First, we investigated the features before edge-etching (Figs. 2a–c). Both $R_{\text{loc}}$ and $R_{\text{nl}}$ values at the CNP increase as the band gap of BLG opens with increasing $|D|$. As shown in Figs. 2b and c, $R_{\text{nl}}$ is an order of magnitude smaller than $R_{\text{loc}}$ for a wide range of $n$ and $D$. Thus, we suppose two possible contributions to $R_{\text{nl}}$, namely, the valley Hall effect and the ohmic contribution. When the displacement field breaks the spatial inversion symmetry and opens the band gap of BLG, a Berry curvature of different signs but the same magnitude is expected to emerge for different valleys near the CNP[6]. Because the Berry curvature acts as an out-of-plane magnetic field in momentum space, the direction of which is opposite for different valleys, a valley current is generated orthogonal to the charge current. This phenomenon is called the valley Hall effect and contributes to $R_{\text{nl}}$. Such phenomena have been studied in monolayer graphene-aligned hBN layers[27, 39] and gapped BLG with dual gates[23, 24] using $R_{\text{nl}}$ measurements, as well as in transition metal dichalcogenides (TMDs), which exhibit valley-dependent optical properties[40-42]. The ohmic contribution arises when part of the bias current, which exponentially decays with the distance from the source terminals, reaches the nonlocal voltage terminals, causing a voltage drop. According to the van der Paw formula[21, 27], the ohmic contribution can be expressed as $R_{\text{Ohmic}} \approx (4/\pi)\rho_{\text{xx}}\exp(-\pi L/W)$, where $\rho_{\text{xx}}$ is the local resistivity, $L$ is the distance from current source terminal 2 to nonlocal voltage terminal 3, and $W$ is the width of the device; in this study, $L$ is taken as 8.2 μm. The $W$ values before and after edge-etching are 8.5 and 6.8 μm, respectively. The dependence of $R_{\text{nl}}$ measured before edge-etching on $n$ and $D$ are well explained by the ohmic contribution ($R_{\text{nl}}/R_{\text{Ohmic}}{\sim}1$) for $|D| < 0.4$ V/nm, without considering the topological valley Hall effect. The deviation of $R_{\text{nl}}/R_{\text{Ohmic}}$ from 1 at $|D| > 0.4$ V/nm can be attributed to the intrinsic edge disorders, which will be discussed later.

Figures 2d–f show the transport properties measured after edge-etching. As shown in Fig. 2e, $R_{\text{loc}}$ at the CNP with $D = -0.61$ V/nm decreases by a factor of 2. Most strikingly, the ratio $R_{\text{nl}}/R_{\text{Ohmic}}$ reaches up to 800 at the highest $D = 0.72$ V/nm (Fig. 2g). This means that $R_{\text{nl}}$ measured in the device with etched edges cannot be explained solely by ohmic contributions. The valley Hall effect, which is a bulk effect, cannot explain this observation neither as the BLG bulk has been protected by the hBN and graphite gate layers and no change of BLG bulk after edge-etching is expected. To explain these drastic changes in $R_{\text{loc}}$ and $R_{\text{nl}}$ after edge-etching, we suggest that the edge-etching process gives rise to edge conduction channels[34, 43-47]. If edge conduction channels are present with the insulating BLG bulk, the bias current from terminals 2 to 6 can flow through these channels in both the counterclockwise (from terminal 2 to terminals 1, S, 5, and then 6) and clockwise (from terminal 2 to terminals 3, 4, D, 7, and then 6) directions. Subsequently, this can give rise to a voltage drop between voltage terminals 3 and 7, and result in the enhancement in $R_{\text{nl}}$ (see schematics in the Supporting Information, Fig. S1b). The formation of edge conduction channels could also explain the observed decrease in $R_{\text{loc}}$ after edge-etching. $R_{\text{loc}}$ at the CNP before edge-etching (blue circles in Fig. 2c) shows

exponential growth with increasing $|D|$, as expected for clean BLG bulk without mid-gap impurity states. $R_{\mathrm{loc}}$ at the CNP after edge-etching (blue circles in Fig. 2f) exhibits saturation, rather than exponential growth, as $|D|$ increases, which is well explained by the edge conduction channels dominating the conduction when the bulk is insulating. Note that $R_{\mathrm{loc}}$ at the CNP near $D = 0$ does not change after edge-etching because bulk conduction dominates over the edge conduction.

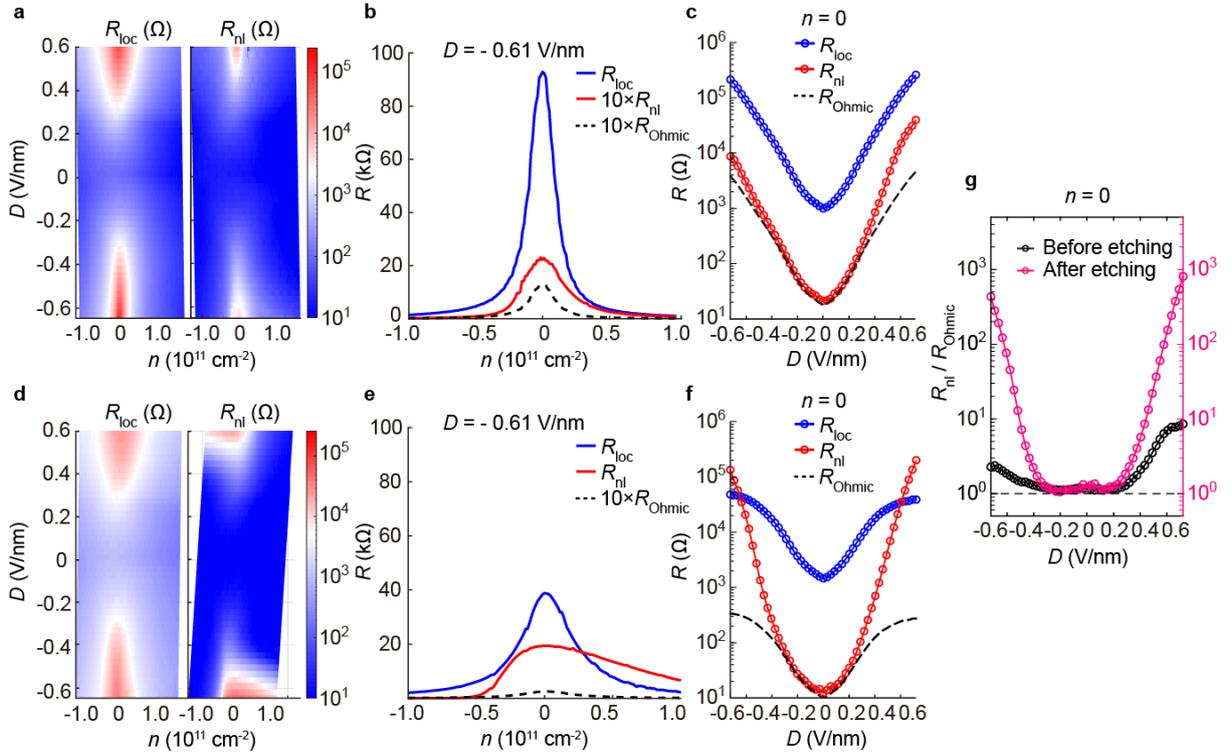

**Figure 2**. Local ($R_{\mathrm{SD,23}}$) and nonlocal resistance ($R_{26,37}$) of BLG device 1 before and after edge-etching. (a) Color-coded plots of the $R_{\mathrm{SD,23}}$ (= $R_{\mathrm{loc}}$) and $R_{26,37}$ (= $R_{\mathrm{nl}}$) of the BLG device before edge-etching as a function the carrier density $n$ and vertical displacement field $D$ at a temperature of 50 K. Color codes are in log scale. (b) Line-cuts of $R_{\mathrm{SD,23}}$ and $R_{26,37}$ taken from (a) at $D = -0.61$ V/nm. The dashed line indicates the expected nonlocal resistance from ohmic contribution ($R_{Ohmic}$) (c) $R_{\mathrm{SD,23}}$, $R_{26,37}$, and $R_{\mathrm{Ohmic}}$ at the CNP ($n =$

0) with respect to $D$. (d-f) Corresponding results of device 1 after edge-etching. (g) The ratio of $R_{nl}$ to $R_{Ohmic}$ before and after edge-etching at the CNP ($n = 0$) with respect to $D$. The line where $R_{nl} / R_{Ohmic} = 1$ is indicated by the black dash.

The presence of unintended edge conducting channels after edge-etching inevitably leads to the underestimation of the transport band gap ($E_g$) in BLG. Figure 3a shows the dependence of $R_{loc}$ on the reciprocal of the temperature ($1/T$) at the CNP of device 1 measured at a fixed $D = 0.4$ V/nm before and after edge-etching. The $E_g$ extracted by fitting the data to the Arrhenius relationship, $R_{loc} \propto \exp(E_g/2k_BT)$, decreases from 37.3 meV before edge-etching to 22.1 meV after edge-etching. Figure 3b summarizes the dependence of the $E_g$ estimated for various devices on $D$ (Supporting Information, Figure S1a) before and after edge-etching. The $E_g$ values estimated for the devices with naturally cleaved edges and those in Corbino geometry without any edges (denoted by circles, Supporting Information, Fig. S2a-c) closely matched the theoretical expectation[9], whereas the $E_g$ values estimated from the device 1 with etched edges (denoted by triangles) are smaller than the theoretical expectations by about 40 %. This suggests that the estimation of $E_g$ needs to be done carefully, taking into account the presence of edge conducting channels.

We discuss several origins of the edge conducting channel. The most trivial origin is the disorder and impurity near the edges of the BLG. Previous imaging studies[44, 47-49] have directly shown the energy dissipation due to impurity/disorder-induced localized states near the etched edges. When such localized states overlap each other, they can support the trivial edge conduction and results in a great enhancement of nonlocal resistance and decrease of local resistance. In fact, the energy dissipation occurs also in natural edges[47] due to the intrinsic disorder, which can explain the deviation the nonlocal resistance from the ohmic contribution in our natural edge BLG device (Figure 2c, $|D| > 0.4$). Earlier investigations[50, 51] have studied the valley-preserved edge channels that form irrespective of the degree of edge disorder, but their length scale of tens of nanometers is much shorter than that of a few micrometers in Hall-bar devices in this study. Thus, such valley-preserved edge channels cannot contribute to the edge conduction observed in our experiment. Another origin of the edge conduction is the charge accumulation due to stray electric field near the edges (Figure 1d). Indeed, our simulations (Supporting Information, Figure S3) demonstrate that the stray electric field induces carriers at the device edge, thereby contributing to the observed decrease in local resistance after edge-etching (Figure 1d). However, our simulation model does not take into account the inhomogeneous charged impurities near the edge, and further investigations are necessary to clarify the origin of edge conducting channels.

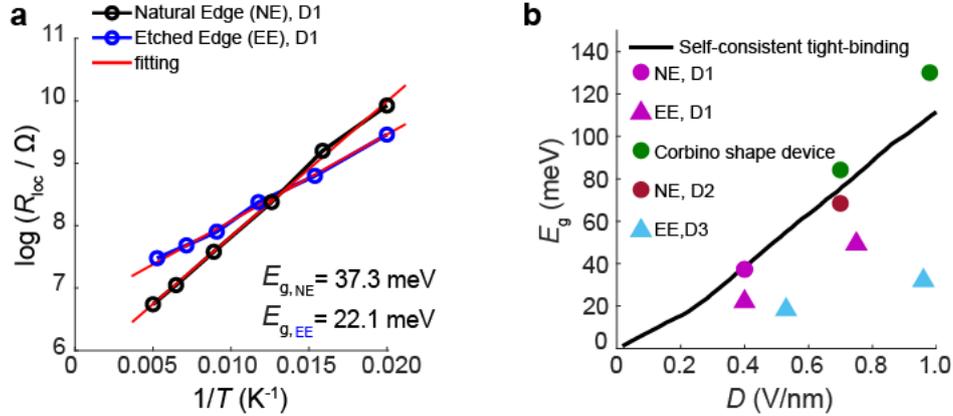

**Figure 3.** Band-gap estimation from the temperature ($T$) dependence of $R_{loc}$. (a) Arrhenius plot of the $T$ dependence of the conductance at the CNP of device 1 when $D = 0.4$ V/nm before and after edge-etching. The black circles and red lines represent the measured data and corresponding exponential fitting, respectively. (b) Band gaps of various devices estimated from the Arrhenius fitting as a function of $D$. The circles represent devices with either natural edges or no edges in Corbino geometry. The triangles represent the device with etched edges. The solid line represents the theoretically calculated band gap[9]. The error bar in the band gap is smaller than the size of the symbols.

CONCLUSIONS

In summary, we investigated the effect of edge-etching on local and nonlocal transport in dual-gated BLG devices. High-quality dual-gated BLG devices with naturally cleaved edges exhibited nonlocal resistance, which can be explained by the trivial ohmic contribution. However, after edge-etching, the nonlocal resistance increased by two orders of magnitude; this finding suggests the formation of additional conduction channels near the etched edges, which can also explain the change in local resistance. Although we cannot disprove the valley Hall effect arising in gapped BLG systems, the nonlocal signals measured in BLG devices with etched edges cannot be attributed solely to the valley Hall effect. In addition, the presence of edge conduction channels can explain why the transport band gap estimated in the edge-etched BLG is smaller than the theoretical value.

METHODS

**Sample preparation and characterizations**: The h-BN, BLG, graphite flakes were exfoliated onto Si/SiO$_2$ (285 nm). All stacks used in this experiment were dry-transferred by using polymer (Elvacite 2552C) stamps and annealed in vacuum at 500 °C for 2 hours to remove bubbles of residue and surface impurities. We identified clean and bubble-free areas within the stacks by using AFM (XE7, Park Systems) topography imaging. The thicknesses of the top and bottom h-BN layers were also measured by AFM.

**Device fabrication and electrical measurements**: The electrodes and etching masks were patterned by standard electron-beam lithography. The h-BN and graphite (BLG) were selectively etched with a reactive ion etcher under $CF_4$ and $O_2$ plasma, respectively. Before we deposit the metal electrodes on the BLG, the top h-BN and graphite gate were etched wider than the contact electrodes to prevent the shortage between the BLG and the top graphite gate. Cr (5 nm) and Au (100 nm) layers were deposited with electron-beam evaporator for the contact electrodes. We also insulated between the bottom graphite gate and the BLG with overdosed cross-linked polymethyl methacrylate (PMMA). Electrical measurements were conducted using a nanovoltmeter (Keithley 2182A) in combination with a current source (Keithley 6221), enabling low-noise delta mode measurements. To measure high resistances in our Corbino-shaped device, we also used a DC setup with a voltage source (Yokogawa 7651) and a current preamplifier (ITHACO 1211).

AUTHOR INFORMATION

**Corresponding Author**

Correspondence to Gil-Ho Lee (lghman@postech.ac.kr)

**Author Contributions**

G.-H.L. and H.-W.J. conceived and supervised the project. H.-W.J. designed and fabricated the devices, performed the measurements together with S.J., S.P., and carried out numerical simulations. T.T. and K.W. provided the hBN crystal. H.-W.J. and G.-H.L. performed the data analysis and wrote the paper.

ACKNOWLEDGMENT

This work was supported by National Research Foundation (NRF) grant (Nos. 2022M3H4A1A04074153, 2022R1A6A3A01086903) and ITRC program (IITP-2022-RS-2022-00164799) funded by the Ministry of Science and ICT. Additionally, support was received from Samsung Electronics Co., Ltd. (IO201207-07801-01). K.W. and T.T. acknowledge support from the JSPS KAKENHI (Grant Numbers 21H05233 and 23H02052) and World Premier International Research Center Initiative (WPI), MEXT, Japan. S.P. was supported by So-Seon Na-Num fellowship, POSTECH Postechian fellowship, and POSTECH Alchemist fellowship.

# Supporting Information for

# Edge dependence of non-local transport in gapped bilayer graphene

*Hyeon-Woo Jeong[1], Seong Jang[1], Sein Park[1], Kenji Watanabe[2], Takashi Taniguchi[3],*

*, and Gil-Ho Lee[1,*]*

[1]Department of Physics, POSTECH, Pohang, Republic Korea

[2]National Institute for Materials Science, Namiki 1-1, Tsukuba, Ibaraki 305-0044, Japan

[3]International Center for Materials Nanoarchitectonics, National Institute for Materials Science, Tsukuba, Japan

[*]Correspondence to Gil-Ho Lee (lghman@postech.ac.kr)

# Supplementary figures

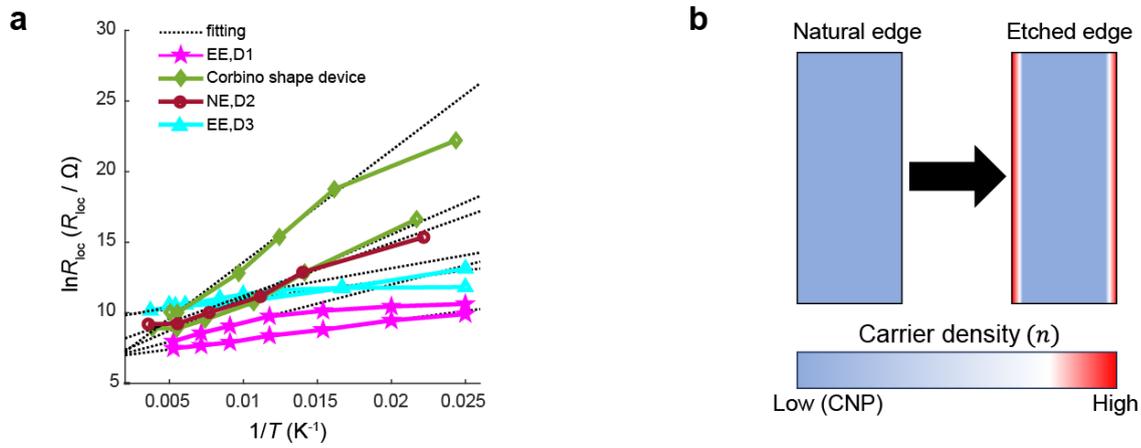

**Figure S4. Temperature dependence data for the devices in Fig. 3B and schematics for carrier doping at the device edge.** (A) Temperature dependence data for the energy gaps plotted in Fig. 3B of the main text. (B) Schematics of edge carrier doping due to stray electric field after edge-etching.

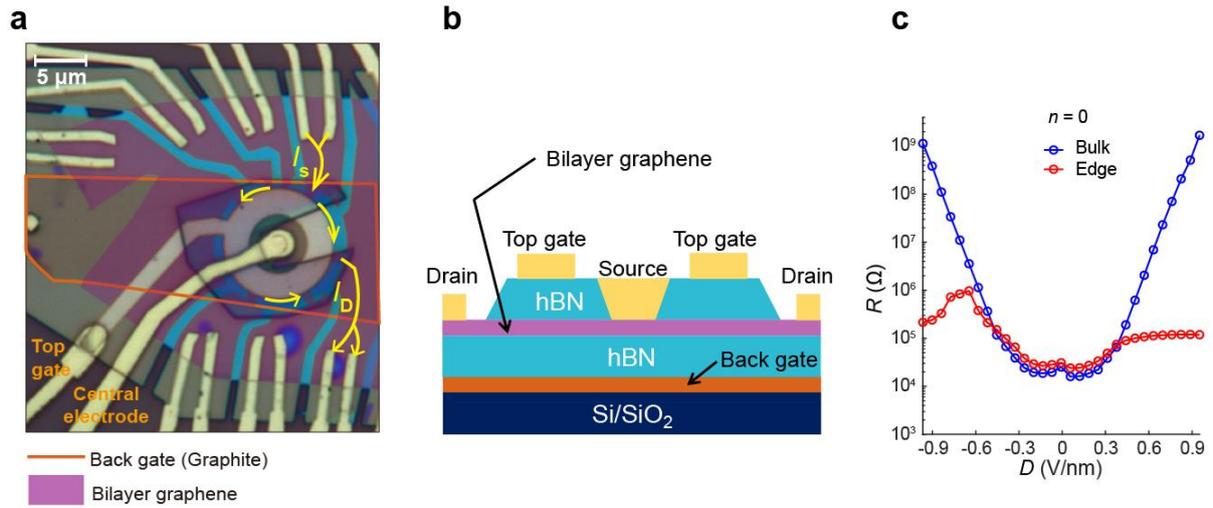

**Figure S2. Images for the Corbino shaped device and the charge-neutral resistance plot.** (a) Optical microscope image of the Corbino-shaped device. The yellow arrows indicate the direction of the bias current flowing from one of outer contact to another outer contact along the etched outer edge of bilayer graphene. (b) Side-view schematics of the device in (a) when the bias current is flowing from central contact to outer contact through the bilayer graphene bulk. (c) Charge-neutral resistance with respect to displacement field, $D$, when the bias current flows through the device bulk (blue) or edge (red).

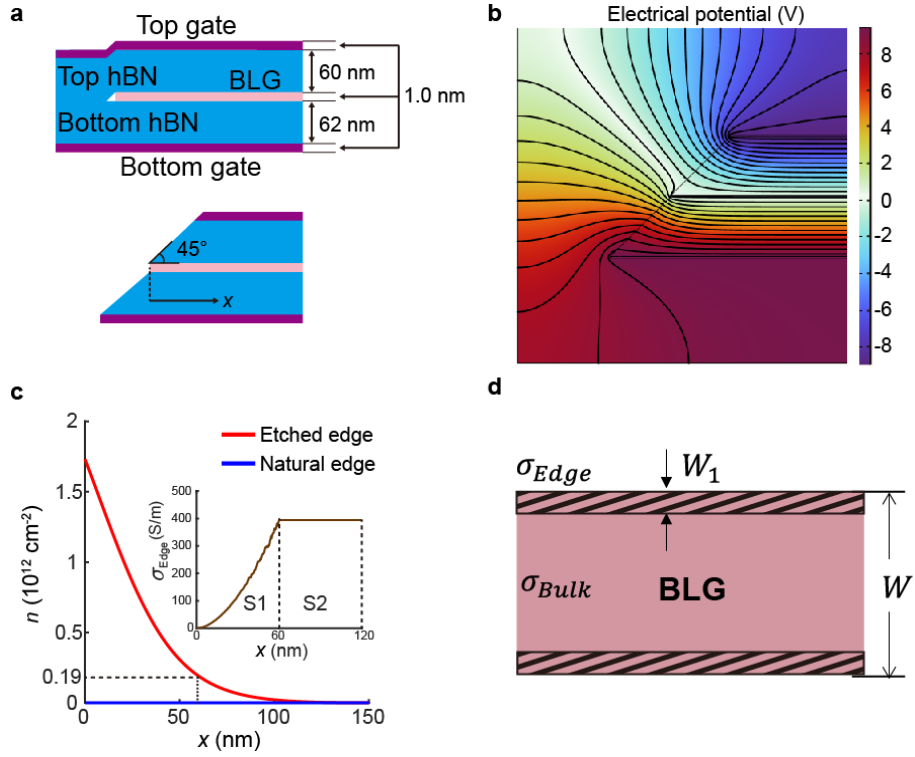

**Figure S5. Stray electric field driven carrier density as a function of the distance from the device edge calculated through the finite element method using COMSOL.** (a) Schematics illustrating the geometry of both the natural edge (upper) and the etched edge (lower) used in the simulations. The dashed black line indicates where the edge starts. (b) Equipotential contour plot from the simulations ranges from $V_{\mathrm{BG}} = 9.4\ V$ to $V_{\mathrm{TG}} = -9.0\ V$ with the equal voltage step. (c) Carrier density calculated as a function of the distance from the edge while the bulk is charge neutral. The displacement field was fixed at $D$=0. 60 V/nm in the calculation. The inset shows the two integral region S1 and S2 to estimate the edge conductance $G_{\mathrm{Edge}}$. (d) Schematic diagram of the physical width ($W$) of the device and anomalously doped width (hatched box, $W_1$) due to the stray electric field.

We used commercial software package (COMSOL simulator) to investigate the charge accumulation induced by the nonuniform electric field at the edge of the device, varying with the distance ($x$) from the edge. In the simulation results, the carrier density ($n$) near the edge ($x < W_1$) showed the order of $10^{12}$ cm$^{-2}$ difference depending on the edge geometry (Figure S3c). $W_1$ in Fig. S3d is a specific length defined with the carrier density exceeding $5.0 \times 10^9$ cm$^{-2}$, resulting in a local resistance ($R_{loc}$) decrease of 10% compared to its charge-neutral resistance. Based on the calculated carrier density, a simple model is used to explain the decreased local resistance after the edge-etching (Fig.1d in the main text). We assumed that the bulk of the device remains unaffected by external contamination before and after edge-etching process, as it is protected by the graphite gates and h-BN, resulting in constant conductivity.

With this assumption, the constant bulk conductivity ($\sigma_{Bulk}$) of the device can be represented as $\sigma_{Bulk} = G_{loc}/W_N$ where $G_{loc}$ is the measured charge neutral conductance at $D = 0.6$ V/nm and $W_N = 8.5$ µm is the width of the device before etching. On the other side, the edge conductivity ($\sigma_{Edge}$) varies with the carrier density $n$ which is a function of $x$ ($x < W_1$). By correlating $G_{loc}(n)$ data from the dual gate measurements with the $x$-dependent carrier density $n(x)$ (Fig. S3c), we can denote the edge conductivity as $\sigma_{Edge}(x) = G_{loc}(x)/W_N$. Consequently, the general form of the conductance from the edge ($G_{Edge}$) is the summation of the infinitesimal conductance $G_{Edge} = 2 \cdot \int_0^{W_1} \sigma_{Edge}(x) \cdot \mathrm{d}x$ for both edges. We divided the integral into two parts at x=60 nm, S1 and S2, under the assumption that the conductivity remains constant within the S2 region, given the maximum carrier doping ($1.9 \times 10^{11}$ cm$^{-2}$) in our experiment. Therefore, the total conductance of the device after edge etching ($G_T$) can be described by $G_T = (W_E - 2 \cdot W_1) \cdot \sigma_{Bulk} + G_{Edge}$. After we substitute the values from the device corresponding to each parameter as follows, $W_E = 6.8$ µm, $W_1 = 120$ nm, $\sigma_{Bulk} = 1.2$ S/m, $G_{Edge} = 48$ µS, then $G_T = 56$ µS ( $R_T = 1/G_T \approx 18$ kΩ ) which is of the same order as the measured conductance after edge etching. Our basic model does not account for the charged impurities at the device edge, which contribute to unpredictable stray electric fields. Nonetheless, the model closely predicts the reduction in measured local resistance following edge-etching, demonstrating that nonlocal transport can occur through the trivial metallic edge channel without a topological origin.